\documentclass[conference]{IEEEtran}
\IEEEoverridecommandlockouts
\usepackage{cite}
\usepackage{amsmath,amssymb,amsfonts}
\usepackage{algorithmic}
\usepackage{graphicx}
\usepackage{textcomp}
\usepackage{xcolor}
\usepackage{svg}
\usepackage{soul}
\usepackage{tabularx}
\usepackage{graphicx}
\usepackage{subfigure}
\usepackage{url}
\usepackage{siunitx}
\usepackage[section]{placeins}
\usepackage{filecontents}

\newcommand{\system}{{TunnelSense}}
\title{Bibliography management: \texttt{biblatex} package}

\newcommand{\fakeparagraph}[1]{\vspace{.5mm}\noindent\textbf{#1.}}
\newcommand{\fakepar}[1]{\fakeparagraph{#1}}
\def\BibTeX{{\rm B\kern-.05em{\sc i\kern-.025em b}\kern-.08em
    T\kern-.1667em\lower.7ex\hbox{E}\kern-.125emX}}
\begin{document}

\title{\system: Low-power, Non-Contact Sensing using Tunnel Diodes\\
}

\author{\IEEEauthorblockN{Lim Chang Quan Thaddeus\IEEEauthorrefmark{1}\textsuperscript{\textsection},
C. Rajashekar Reddy\IEEEauthorrefmark{1}\textsuperscript{\textsection},
Yuvraj Singh Bhadauria\IEEEauthorrefmark{2},\\ 
Dhairya Shah\IEEEauthorrefmark{1}\textsuperscript{\textsection},
Manoj Gulati\IEEEauthorrefmark{1},
Ambuj Varshney\IEEEauthorrefmark{1}}
\IEEEauthorblockA{\IEEEauthorrefmark{1}National University of Singapore\space
\\
Email: ambujv@nus.edu.sg}}


\maketitle
\begingroup\renewcommand\thefootnote{\textsection}
\footnotetext{Authors contributed equally to the work.}

\begingroup\renewcommand\thefootnote{}
\footnotetext{\IEEEauthorrefmark{2}Yuvraj Singh Bhadauria conducted this work when he was visiting National University of Singapore from BITS Pilani (Goa campus), India.}
\begin{abstract}
Sensing the motion of physical objects in an environment enables numerous applications, from tracking occupancy in buildings and monitoring vital signs to diagnosing faults in machines. Typically, these application scenarios involve attaching a sensor, such as an accelerometer, to the object of interest, like a wearable device that tracks our steps. However, many of these scenarios require tracking motion in a noncontact manner where the sensor is not in touch with the object. A sensor in such a scenario observes variations in radio, light, acoustic, and infrared fields disturbed by the object's motion. Current noncontact sensing mechanisms often require substantial energy and involve complex processing on sophisticated hardware. We present \system, a novel mechanism that rethinks noncontact sensing using tunnel diode oscillators. They are highly sensitive to changes in their electromagnetic environments. The motion of an object near a tunnel diode oscillator induces corresponding changes in its resonant frequency and thus in the generated radio waves. Additionally, the low-power characteristics of the tunnel diode allow tags designed using them to operate on less than \SI{100}{\micro\watt} of power consumption and with a biasing voltage starting at \SI{70}{\milli\volt}. This enables prolonged tag operation on a small battery or energy harvested from the environment. Among numerous applications enabled by the \system\space system, this work demonstrates its ability to detect breathing at distances up to \SI{30}{\centi\meter} between the subject and the \system\space tag.

\end{abstract}

\section{Introduction}
Sensing the motion of objects enables numerous applications. The ability to track body movements enables gesture control for computing devices~\cite{humansensing,bfvls}. For example, tracking subtle body movements can help detect vital signs~\cite{breatheband2023,vitalradio,bioscatter} such as heart and breathing rates. Additionally, motion sensing can apply to other domains, such as tracking machine vibrations to gather crucial information about their health~\cite{smile}, which can help detect anomalies before breakdowns occur. Generally, the mechanisms for tracking these motions are classified into contact-based and noncontact-based methods.

A contact-based method involves attaching a sensor, like an accelerometer, to the object. We capture sensor data through a microcontroller, process it, and then communicate it using a transceiver. Wearable devices that track activities like walking and sleeping are examples of applications that use contact-based methods. However, contact-based methods can restrict application scenarios severely, as attaching a sensor to an object may not always be feasible or desirable. Consequently, there has been significant interest in the development of noncontact-based methods for motion tracking.

A noncontact-based method involves using a sensor to observe changes in electromagnetic~(light or radio frequency), acoustic, or thermal fields in the environment caused by object motion. Often, the sensor generates these fields and then observes changes in them due to  motion of an object. 

\begin{figure}[t!]
    \centering
    \includegraphics[width=0.42\columnwidth]{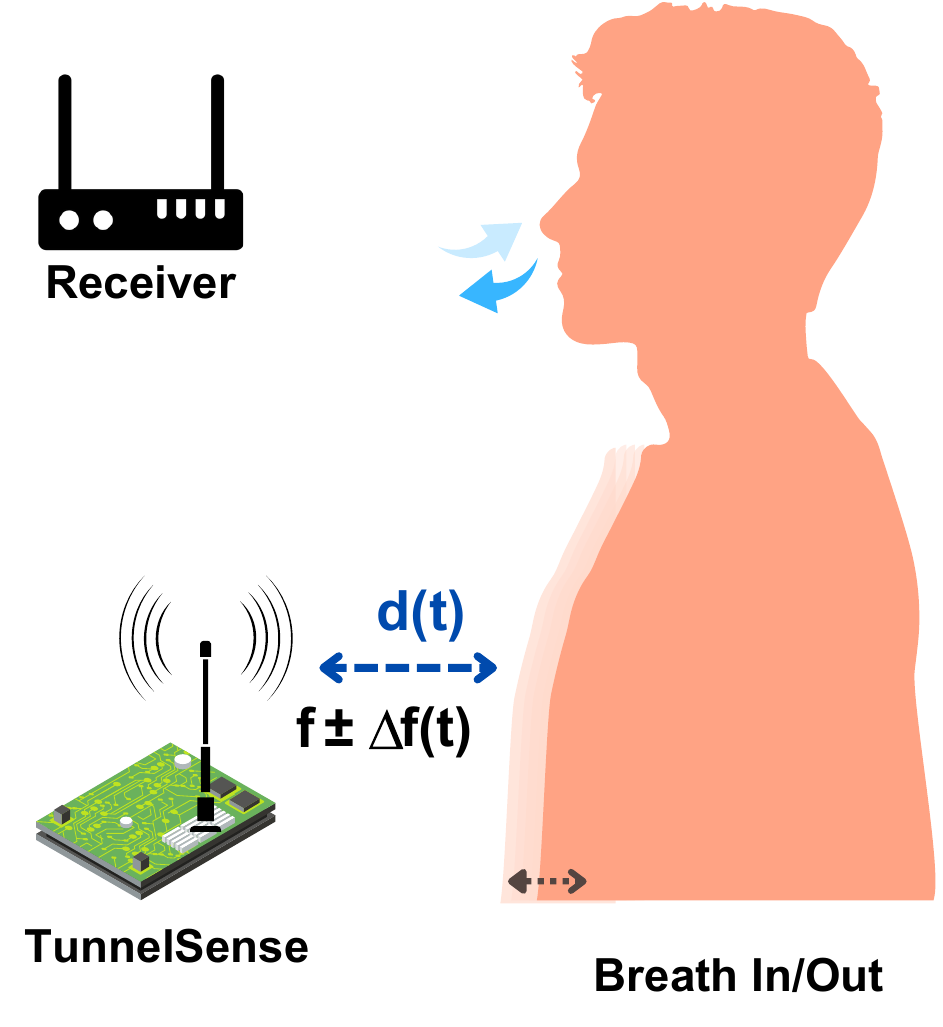}
    \vspace{-4mm}
    \caption{\system\space enables noncontact sensing of the movement of objects in its vicinity. The movement of macroscopic objects causes changes in the electromagnetic environment of the tunnel diode oscillator, causing changes in its resonant frequency.  It operates at under \SI{100}{\micro\watt}s of power consumption, and a commodity transceiver can track these changes in the frequency of the radio waves. In this work, we apply \system\space to sense a person's breathing.}
   \vspace{-6mm}
    \label{fig:TunnelSense}
\end{figure}

Passive infrared sensors track infrared radiation emitted by the human body or other environmental objects. A Fresnel lens concentrates this radiation onto a pyroelectric element. This mechanism can detect motion due to changes in the infrared radiation incident on the pyroelectric element. Similarly, a light sensor tracks changes~\cite{bfvls, humansensing} in light intensity due to obstructions caused by the motion of  objects blocking the light rays. However, these systems require extensive deployment of sensors throughout the environment and are limited to tracking coarse movements near the sensor, thus limiting their effective field of view for sensing. Moreover, most detected events need to be wirelessly communicated, which presents another challenge. Typically, this step requires interfacing the sensor with an embedded device equipped with a  transceiver, further increasing the system's energy consumption~\cite{varshney2022judo,lorea, ambientbackscatter}.

The most widely used mechanism for enabling noncontact-based sensing involves radio frequency (RF) signals. In these systems, a device generates a strong radio wave that interacts with objects in the environment, potentially altering the phase, frequency, or amplitude of the reflected signal. This mechanism has enabled numerous sensing applications~\cite{shah2019rf}, forming the basis of radar, tracking a person's vital signs~\cite{breatheband2023,vitalradio,bioscatter,hui2018monitoring}, and even assessing food quality~\cite{agritera23}.

However, RF-based sensing systems face challenges. Generating a radio wave is an energy-intensive task. Typically, these systems emit a radio wave with the maximum power allowed in the license-free band. The radio wave interacts with objects, and the reflected signal usually has a similar frequency to the incident wave. Therefore, a critical step involves extracting of the weak reflections in the presence of a much stronger radio wave, requiring complex mechanisms such as self-interference cancellation circuitry. Finally, a software-defined radio~(SDR) processes the extracted signal. This architecture demands energy-intensive steps, sophisticated hardware, and complex processing, making it expensive and challenging to deploy such systems at a large scale~\cite{ravichandran2015wibreathe, rahman2015dopplesleep}.

We present a noncontact-based mechanism, \system, that is closely related to RF sensing systems yet has unique attributes that distinguish it from them and overcome their drawbacks. Specifically, \system\space facilitates the generation of radio waves and communicates this motion information with power consumption below \SI{100}{\micro\watt}, even when generating signals in the ISM band, such as at 868 MHz. Furthermore, it can operate at very low voltages, starting from \SI{70}{\milli\volt}, overcoming the high cold-start voltage limits typical of energy harvesters. Importantly, the motion information is embedded in the generated radio wave, eliminating the need to extract weak signals from the stronger incident signal, thus allowing reception on commodity radio transceivers, or low-cost SDRs.

\fakepar{\system\space design} Tunnel diodes are semiconductor devices that exhibit a region of negative resistance. This means that as we increase the bias voltage across it, the current through the diode increases until it reaches a specific voltage, after which it starts to decrease despite further increases in voltage. This phenomenon occurs at low voltages and current consumption. Specifically, the negative resistance region of the tunnel diode~(1N3712) used in this work begins at a peak current consumption of \SI{1}{\milli\ampere} and a bias voltage of \SI{70}{\milli\volt}.

Building on our earlier work, TunnelScatter~\cite{varshney2019tunnelscatter} and Judo~\cite{varshney2022judo}, we designed a tunnel diode oscillator~(TDO). We achieve this by coupling a tunnel diode with a resonant circuit. Specifically, we configure the resonant circuit to enable the generation of radio waves at frequencies of a few hundred MHz, utilizing the commonly available license-free band. The tunnel diode oscillator facilitates the generation of these radio waves with a peak power consumption of under \SI{100}{\micro\watt}s.

\emph{So, how do we use a tunnel diode oscillator for sensing~?} In designing the tunnel diode oscillator for low-power consumption, we accept inevitable trade-offs. The tunnel diode oscillator's phase noise is high, and it is unstable. Varshney et al. have utilized the injection locking mechanism to stabilize the tunnel diode oscillator and design the transmitter~\cite{varshney2022judo}. However, without an external signal for injection locking, the tunnel diode oscillators are highly sensitive to changes in the electromagnetic environment around them. Specifically, the objects nearby behave as part of its resonant circuit, causing shifts in the tunnel diode oscillator's frequency in response to nearby object properties and motion.
 
We employ \system\space to sense a person's breathing rate.  In detecting breathing, when a person sits close to the tunnel diode oscillator, the carrier signal interacts with the person, and the reflected signal returns to the tunnel diode oscillator We illustrate this through the Figure~\ref{fig:TunnelSense}. The frequency of the tunnel diode oscillator changes in response to breathing, creating distinctive patterns in the frequency of the generated radio waves. This allows us to estimate the person's breathing rate. Since this method does not require extracting the reflected signal from the stronger carrier signal, it simplifies the receiver's design. Most commodity receivers can track the signal's frequency, which simplifies processing and reduces the overall system's cost and complexity compared to traditional radio frequency breathing rate sensing systems.

\fakepar{Summary of results} We conduct extensive experiments and highlight the following major findings from our work:
\begin{itemize}

\item \system\space effectively senses a person's breathing rate, with the tag itself placed up to \SI{30}{\centi\meter} from the subject. 

\item  The \system\space tag operates at voltages as low as \SI{70}{\milli\volt} and under \SI{100}{\micro\watt} power consumption. This allows for extended operation of the tag on small batteries or harvested energy. We estimate that the \system\space tag can be powered through energy harvested from a photodiode which is stored on a small capacitor.

\end{itemize}

\section{Background}

\fakepar{Energy harvesting} We are surrounded by energy sources that can power embedded systems, such as thermal gradients, light, and radio waves. A typical energy harvesting device operates as follows: Firstly, an energy harvesting element, such as a solar cell, antenna, or thermoelectric generator (TEG) element, extracts energy from the environment. Secondly, it directs the gathered energy to an energy-harvesting integrated circuit (EHIC). Over time, the EHIC stores harvested energy, typically in a capacitor. Next, after the voltage across the storage element, usually a capacitor, exceeds the cold-start voltage, the EHIC activates additional circuitry, causing a rapid increase in energy absorption. Typically, the cold-start voltage is restricted to \SI{200}{\milli\volt} due to transistors' limits~\cite{swanson}, as Swanson and Meindl dictated. Finally, the accumulated energy powers the embedded device. Two significant thresholds exist: cold-start and the minimum required to operate the device. The cold-start voltage is the most critical, as ambient sources often cannot provide sufficient energy to take the EHIC beyond the cold-start voltage, leaving many sources in the environment untapped~\cite{shrivastava201510}. \system, due to the low power/voltage operation of tunnel diodes, allows operation on such sources.

\fakepar{Tunnel diodes for communication} Recent efforts use tunnel diodes to design low-power communication systems. Amato et al. demonstrate a gain as high as 40 dB using tunnel diode-based reflection amplifiers to design a backscatter tag that operates in the 5 GHz frequency band~\cite{amato2015long,amato2018tunnel2,amato2018tunneling}. Similarly, Varshney et al. design a reflection amplifier for a backscatter tag operating in the 868 MHz frequency band~\cite{varshney2019tunnelscatter} with a gain of 35 dB and demonstrate reception on commodity transceivers. GPSMirror designs a relay to extend the reach of GPS signals to indoor environments~\cite{gpsmirror}. Adeyeye et al. design a low-powered repeater for RFID tags~\cite{Adeyeye}. Eid et al. design \SI{88}{\milli\volt} RFID tags that harvest energy from a carrier signal and use tunnel diodes for both a reflection amplifier and to generate a low-frequency signal for driving the reflection amplifier~\cite{eid}. Beyond reflection amplifiers, efforts exist to design oscillators using tunnel diodes~\cite{varshney2019tunnelscatter}, which enable the design of energy-efficient transmitters~\cite{varshney2022judo} and carrier emitters~\cite{tunnelemitter}. Varshney et al. discuss that tunnel diode oscillators can be unstable~\cite{varshney2022judo}, and hence, they injection lock them using an external carrier signal for reliable transmissions. Nonetheless, the instability of the tunnel diode oscillator is highly desirable for designing noncontact sensing mechanisms. We exploit it in the design of \system\space and demonstrate the mechanism applied for sensing a person's breathing rate.

 \fakepar{Breathing rate}  Initial efforts involved tracking the motion of the torso or other body parts using an accelerometer to estimate breathing rates. This makes them power-consuming, and due to the contact based nature of the mechanism, it is not  convenient for the subject~\cite{accelbr1, accelbr2, accelbr3}. Consequently, efforts have been made to design noncontact systems. Acoustic signals can enable the detection of breathing~\cite{soundsensing, nandakumar2015contactless}. Nonetheless, transmitting and processing acoustic signals is an energy-consuming task. Finally, most noncontact breathing rate systems use radio frequency signals~\cite{wifiwall}. These systems require complex algorithms on sophisticated hardware~\cite{yue2018extracting} and are energy-expensive, making it  challenging to deploy them at a large scale. This is because they require extracting the weak reflected signal off bouncing off the object from the strong ambient signal incident on the target object.  

  Our system is closely related to these sensing systems. It does not require complex designs, as the frequency of the incident carrier signal itself changes and tracks the object undergoing motion. This enables tracking through a simple transceiver. Furthermore, we generate signals at much lower strengths (approx. \SI{-19}{\decibel}m) under \SI{100}{\micro\watt}s. This enables \system\space to operate on minuscule energy harvested from the environment. Nonetheless, we acknowledge that \system\space makes trade-offs and supports sensing only at short distances compared to radio frequency sensing systems.

\fakepar{Radar and RFIDs}  Systems based on the concept of radar have been used to estimate breathing rate. These systems achieve much larger sensing ranges and operate well in noisy and outdoor environments. Nonetheless, they require complex processing to extract the reflected signal bouncing off the chest of the subject from the stronger incident carrier signal. These systems are significantly power-consuming and expensive. Our work is closely related to RFID systems~\cite{tagbreath} for breathing rate estimation. However, RFIDs require energy-expensive, complex, and costly readers that typically generate a strong carrier signal. Our system is much simpler, generates a carrier signal that is orders of magnitude weaker, can use commodity transceivers or low-cost software-defined radios, and thus enables more sustainable and ubiquitous deployments.

\section{Design}
\begin{figure}[t!]
\centering
\includegraphics[width=0.40\columnwidth]{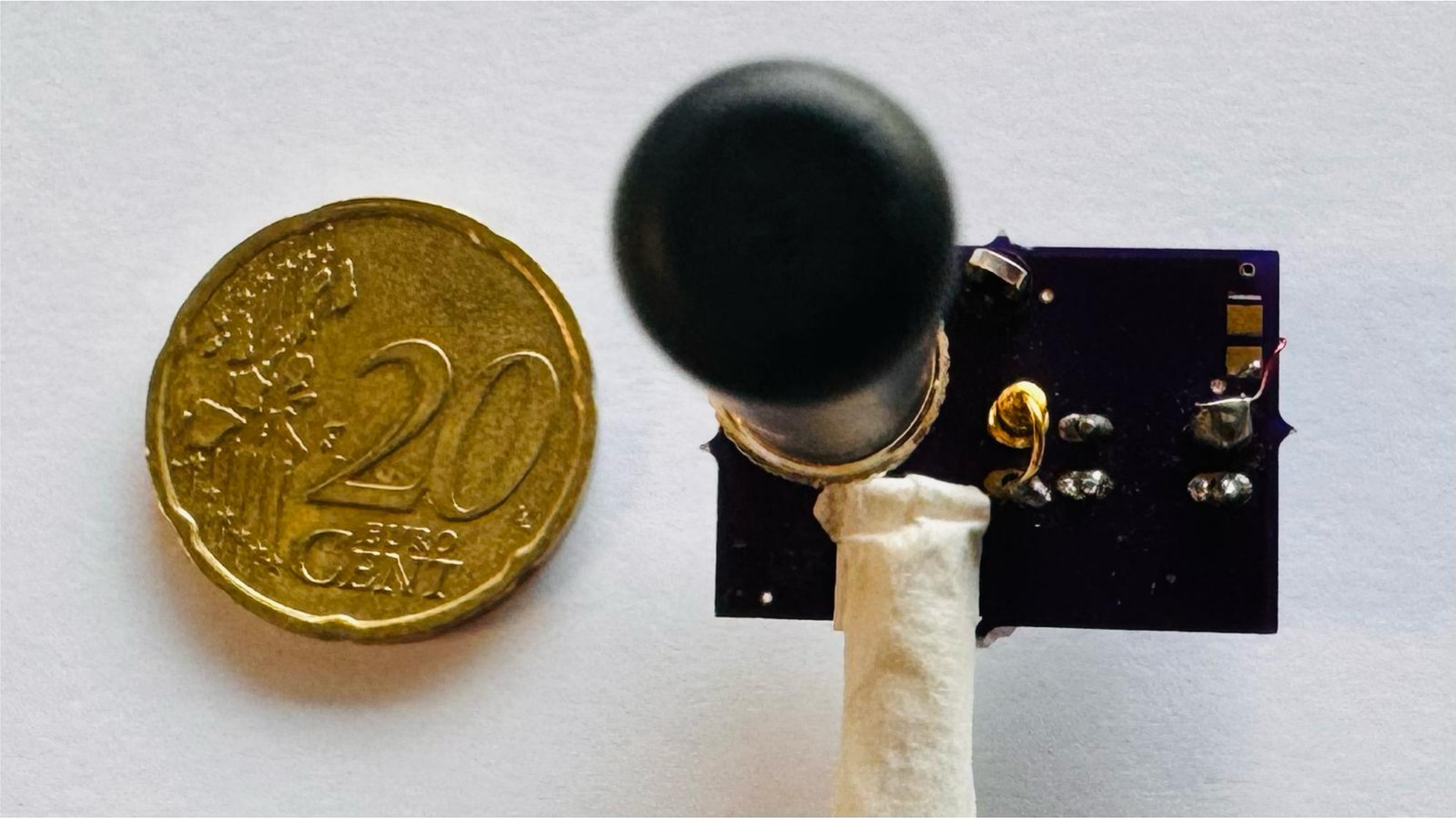}
\vspace{-4mm}
\caption{\system\space is designed with minimal components, which allows the platform in its simplest form to be as compact as a coin cell. The simplicity of the circuit and low-power consumption can also enable the realization of the sensor in novel form factors, such as postage stamp sized stickers.
}
\vspace{-4mm}
\label{fig:hardwareprototype}
\end{figure}
\begin{figure}[t!]
\centering
\includegraphics[width=0.37\textwidth]{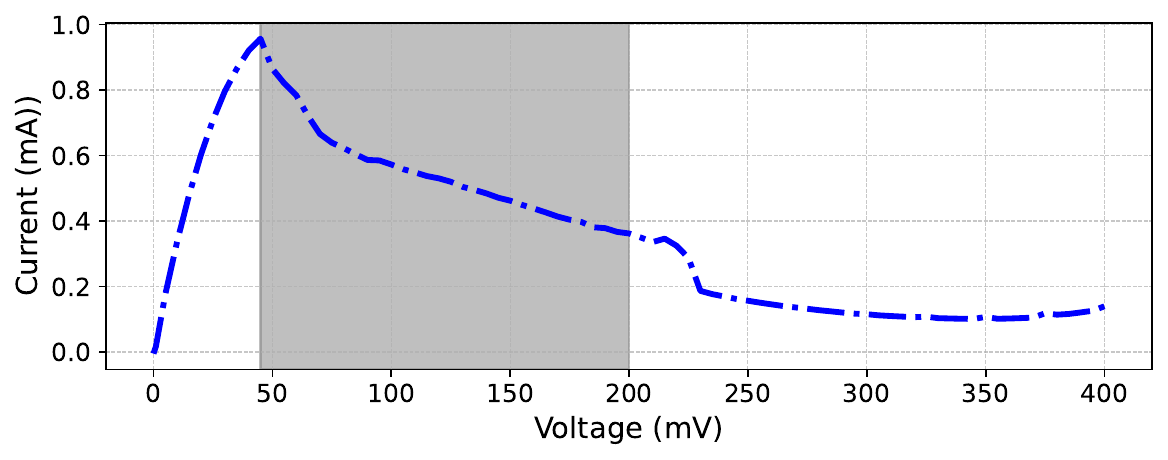}
\vspace{-4mm}
\caption{Tunnel diode shows negative resistance characteristics. Thus, an increase in the bias voltage across the tunnel diode~(1N3712) leads to a drop in the current through the diode (after reaching a threshold voltage). We show a part of this region where \system\space operates with the shaded area.}
\vspace{-4mm}
\label{fig:IV_characteristics}
\vspace{-2mm}
\end{figure}

\system\space comprises a tag (as shown in Figure~\ref{fig:hardwareprototype}), a subject, and one or more receiving devices. The tag utilizes a tunnel diode oscillator to generate radio waves. As the subject moves—marked by the macroscopic movements of the chest and torso during breathing—these movements alter the electromagnetic environment around the tunnel diode oscillator. This change triggers frequency drifts in the radio waves. The radio waves then travel toward a receiver located up to tens of meters away from the tag, which captures and analyzes the frequency drifts to detect the activity causing the motion.

\fakepar{\system\space Tag} The tag can be considered to be broadly organized using two blocks: an energy harvesting circuit and a tunnel diode circuit. The energy harvesting circuit includes a source such as a solar cell, photodiodes, a matched antenna, or a TEG element to capture energy from the surrounding environment. This harvested energy is stored in a storage element, such as a capacitor, which powers the tunnel diode circuitry. Given the low-power and low-voltage characteristics of the tunnel diodes, the energy harvesting circuit might not include an EHIC. In such instances, the energy source directly replenishes the capacitor, particularly when ambient sources fail to provide sufficient voltage to exceed the cold-start voltage of the EHIC. Moreover, the energy harvesting circuit might also incorporate a regulator to maintain a consistent bias voltage for the tunnel diode circuit, as previously demonstrated by Varshney et al.~\cite{tunnelemitter}, who noted that the resonant frequency of tunnel diode oscillators depends on the bias voltage.

We design the tunnel diode circuit by building on TunnelScatter~\cite{varshney2019tunnelscatter} and Judo~\cite{varshney2022judo}. We specifically select the 1N3712 tunnel diode for its low-power characteristics. This tunnel diode operates efficiently in the negative resistance region with a voltage range of \SI{70}{\milli\volt} to \SI{150}{\milli\volt} and a peak biasing current of \SI{1}{\milli\ampere}. Consequently, we can bias it under \SI{100}{\micro\watt} of power consumption. We present the Current (I) and Voltage (V) characteristics of this specific tunnel diode in Figure~\ref{fig:IV_characteristics}. We then couple the tunnel diode to a resonant circuit, which determines the operating frequency along with its properties and the circuit design as shown in Figure~\ref{fig:circuit}. We set up the resonant circuit to enable oscillations in the standard ISM band and configured it to operate at 868 MHz.

\fakepar{Sensing mechanism} Tunnel diode oscillators are noisy, have high phase noise~\cite{varshney2019tunnelscatter}, and are unstable. These tradeoffs are inevitable to allow for low-power consumption. Varshney et al. demonstrate that the frequency of the tunnel diode oscillator drifts with changes in environmental conditions like temperature, humidity, and nearby motion~\cite{varshney2022judo, tunnelemitter}. \system\space exploits the unstable nature of the tunnel diode oscillator for sensing. In particular, the tunnel diode oscillator, due to its unique negative resistance characteristics, is highly susceptible to changes in its electromagnetic environment. The presence of an object in the vicinity can cause changes in the inductive and capacitive values of the resonant circuit. Furthermore, the reflected signal from the object further contributes to these changes. Consequently, the presence of an object in the vicinity and its motion cause changes in the frequency of the generated carrier signal forming basis for \system\space.

We illustrate the sensing mechanism with an experiment. We place objects made of different materials at varied distances from the tunnel diode oscillator. Next, we observe the drift in the frequency of the carrier signal generated by the tunnel diode oscillator. We observe that both the material and position influence the frequency of the radio waves, as illustrated in Figure~\ref{fig:materials}. This observation highlights the tunnel diode oscillator's sensitivity to motion in its vicinity.

\begin{figure}[t!]
\centering
\includegraphics[width=0.44\textwidth]{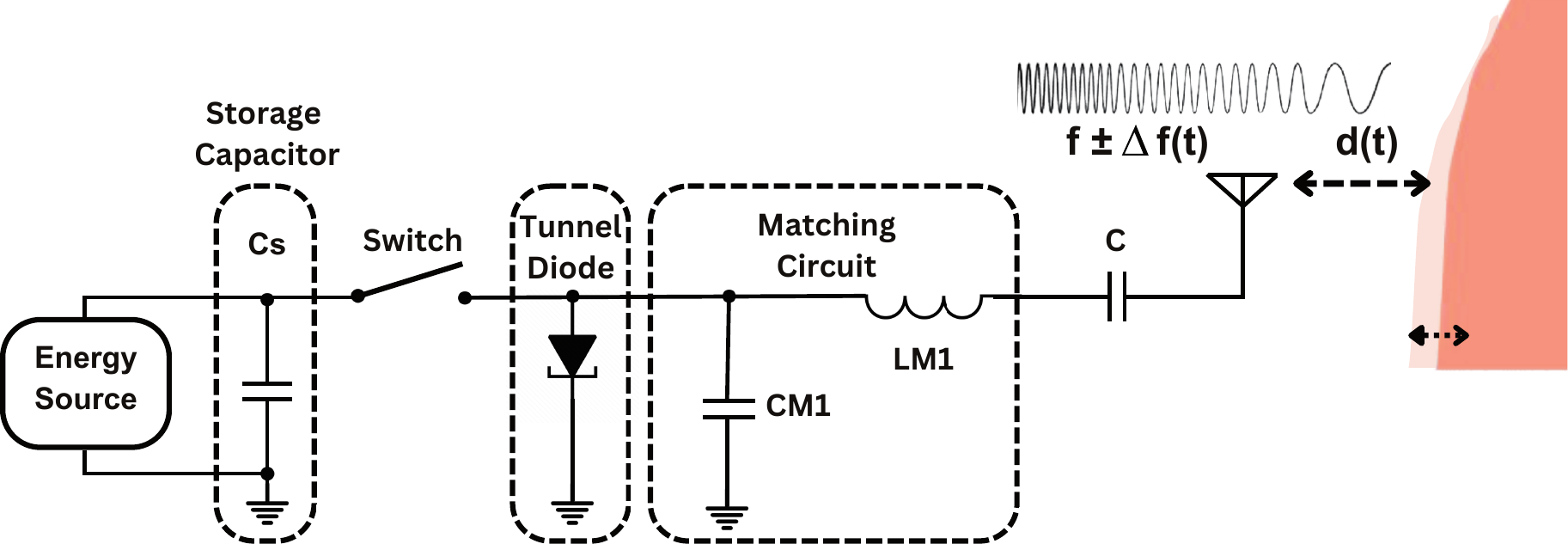}
\vspace{-4mm}
\caption{The tag consists of a tunnel diode oscillator integrated with a resonant circuit. It is configured to generate a carrier signal at the \SI{868}{\mega\hertz} band. When motion occurs near the tunnel diode, the reflected signal propagates back to the oscillator. Additionally, the nearby object induces electrical and magnetic interactions with the circuit, resulting in frequency drifts in the generated carrier signal that correlate with the object's motion.}
\label{fig:circuit}
\vspace{-4mm}
\end{figure}

\begin{figure}[t!]
    \centering
    \includegraphics[width=0.4\textwidth]{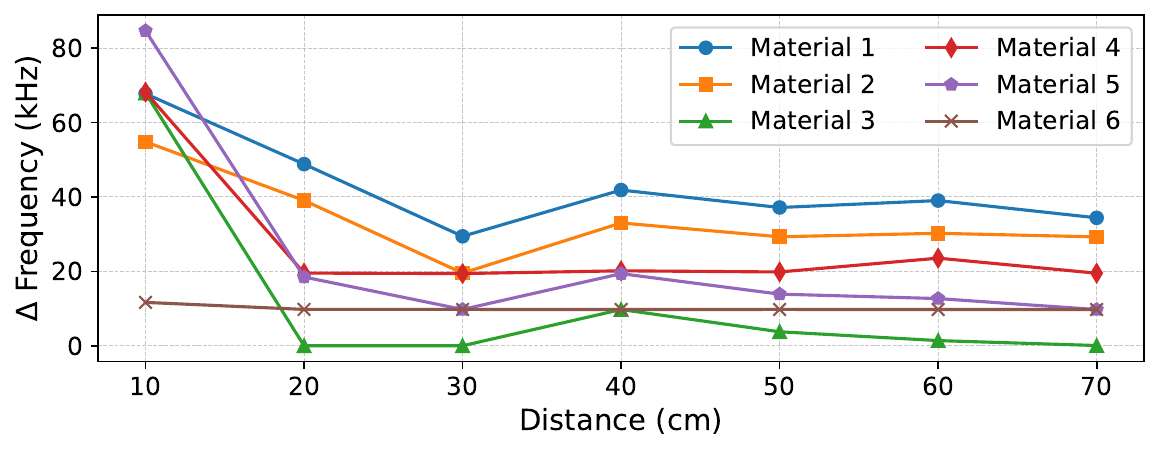}
    \vspace{-4mm}
    \caption{The choice of material and location of object from  the tag impacts the frequency drift of the radio waves generated by the tunnel diode oscillator.}
    \label{fig:materials}
    \vspace{-6mm}
\end{figure}
\fakepar{Receiving device} Movements near the tunnel diode oscillator result in a frequency drift of the generated carrier signal. Unlike conventional radio frequency sensing systems, the receiver does not need to isolate weak reflections from the strong carrier signal. This frequency drift in response to the motion of macroscopic objects simplifies the receiver's design. Commodity radio transceivers, which allow for easy adjustment of the operating frequency and estimation of signal strength, can adapt to track the frequency generated by the tunnel diode oscillator. This enables commodity devices deployed in the environment to detect motion. It also supports using simpler and more affordable software-defined radios, such as TinySDR~\cite{tinysdr} and Adalm Pluto~\cite{adalm_pluto}, as receivers, thereby lowering the cost and complexity of deployment.

\section{Evaluation}
We evaluate the ability of \system\space  to sense the subject's breathing rate. We conduct experiments in a complex radio propagation environment that includes the inside of a university office (with static and moving people) and an outdoor setting. Our main findings indicate that we can track the breathing rate from up to 30 cm away from the tag.

\fakepar{Setup} We illustrate the setup for the experiments in this section in Figure~\ref{fig:setup}. We power the \system\space tag with an external battery for these experiments to facilitate controlled testing. However, the tag's low power consumption allows operation using harvested energy without batteries. We place the sensor close to a subject seated on a chair. The subject wears a Vernier GoDirect~\cite{godirect_resp_belt} respiration belt across the chest, serving as the ground truth for our experiments. This belt employs a force sensor with an adjustable nylon strap around the chest to measure respiration effort and rate. We calculate the breathing rate by analyzing the changes in force sensor values. We set the tunnel diode oscillator to emit radio waves at 868 MHz, radiating the signal through an omnidirectional antenna. We track the frequency changes of the radio waves emitted by the system tag using a simple, low-cost software-defined radio, the ADALM-PLUTO evaluation kit~\cite{adalm_pluto}.
\begin{figure}[t!]
    \centering
    \includegraphics[width=0.42\textwidth]{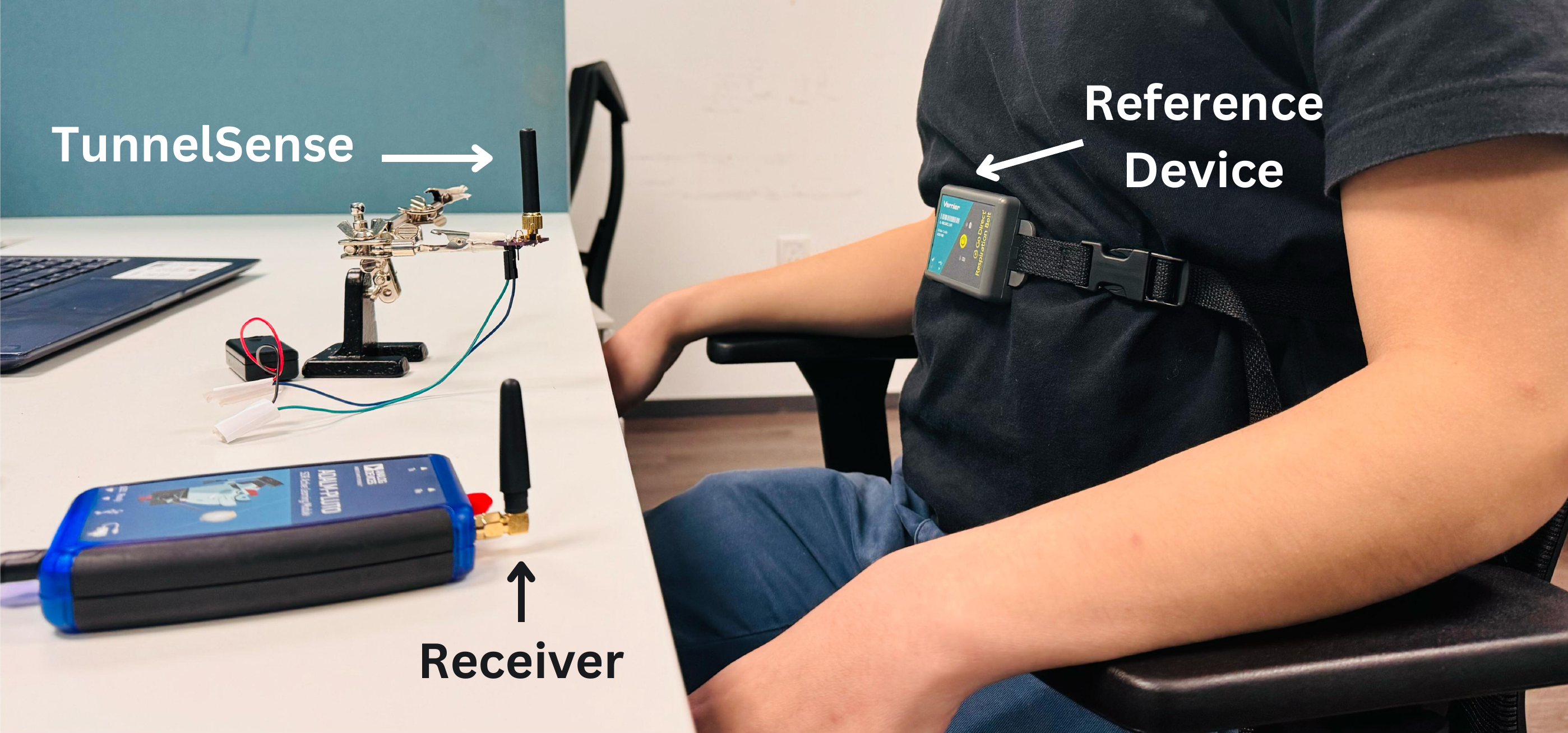}
    \vspace{-3mm}
    \caption{\system\space and the receiver are located close to the subject. The subject wears a breathing monitoring device to collect ground-truth data.}
    \label{fig:setup}
    \vspace{-4mm}
\end{figure}

\begin{figure}[t!]
    \centering
    \includegraphics[width=0.42\textwidth]{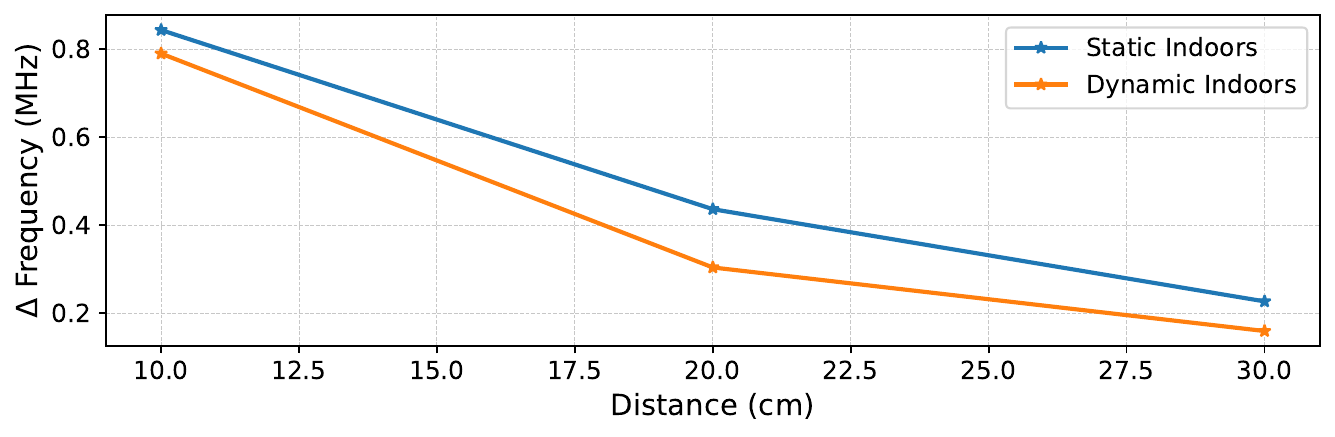}
    \vspace{-4mm}
    \caption{Varying distance between subject and tag causes drift in radio wave frequency to decrease at larger distance between subject and tag. }
    \label{fig:delta_freq_vs_dist}
    \vspace{-6mm}
\end{figure}

\fakepar{Distance from tag} We investigate and measure the magnitude of frequency drift of radio waves generated by the tunnel diode oscillator as we vary the distance between the tag and the subject's chest. We position the subject at various distances from the tag, and Figure~\ref{fig:delta_freq_vs_dist} shows the drift magnitude at these distances. We observe that the frequency drift decreases as the distance from the tag increases. Notably, the drift magnitude becomes small at a distance of \SI{30}{\centi\meter} which prevents our receiver from distinguishing the variations and limits the maximum sensing range of the system. Nonetheless, a more sensitive and capable software-defined radio or higher-gain antenna may allow us to distinguish these changes and help increase the sensing range.
\begin{figure}[t!]
    \centering
    \includegraphics[width=0.42\textwidth]{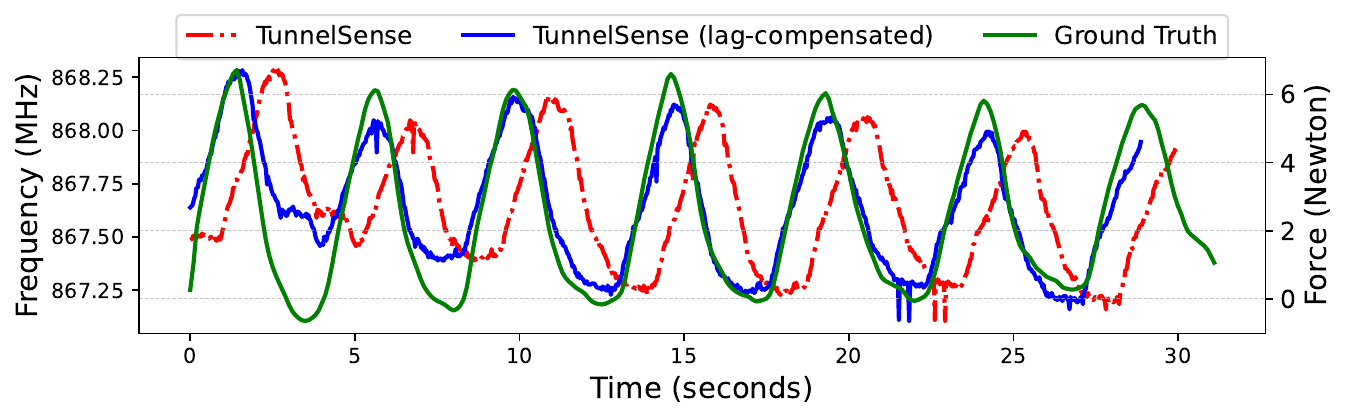}
    \vspace{-4mm}
    \caption{\system\space senses breathing of a person in an indoor (static) environment. The  frequency drifts in the tunnel diode oscillator closely tracks the ground truth. The tag was located at a distance of \SI{10}{\centi\meter} from the subject.}
    \label{fig:breathing_comparison_quiet_10cm}
    \vspace{-4mm}
\end{figure}

\fakepar{Indoor static environment} We evaluate \system\space for monitoring breathing rates in an indoor environment isolated from external disturbances~(no person walking in the vicinity). We conducted experiments with the subject seated inside a university office, placing the system at varying distances from the chest. This setting, characterized by minimal disturbance, closely represents an office environment without interference. It also allows us to assess the system's performance in a static environment. We show the results of this experiment in Figure~\ref{fig:breathing_comparison_quiet_10cm}. It illustrates the frequency variations of the tunnel diode oscillator, which closely match the variations from the raw force values recorded by the reference device. We encountered delays in aligning data from the reference device with \system, primarily due to the acquisition and computation processes on the receiver. We address these delays by calculating the cross-correlation with the reference data and compensating for the lag as shown in Figure~\ref{fig:breathing_comparison_quiet_10cm}.
\begin{figure}[t!]
    \centering
    \includegraphics[width=0.42\textwidth]{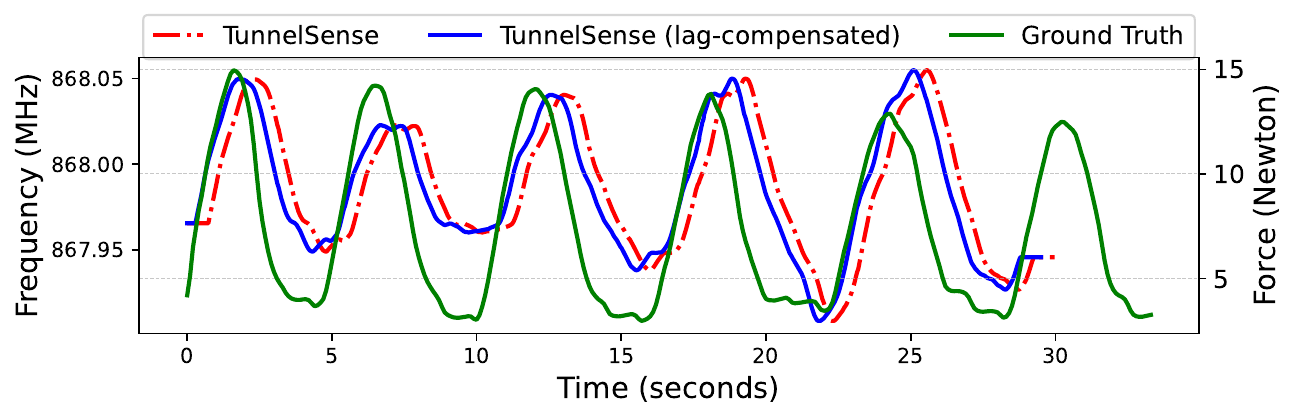}
    \vspace{-4mm}
    \caption{\system\space senses the breathing of a person even in an indoor (dynamic) environment. There are people walking close to the subject, yet patterns corresponding to the breathing  are observed in the frequency drifts of the radio waves. The tag was positioned 30 cm away from the subject.}
    \label{fig:breathing_comparison_lab_30cm}
    \vspace{-7mm}
\end{figure}

\fakepar{Indoor dynamic environment} Next, we evaluate the system in a dynamic indoor environment, where people walk near the subject. We present the results of these experiments in Figure~\ref{fig:breathing_comparison_lab_30cm}. Despite additional disturbances in the environment, the system successfully tracks the breathing rate, closely aligning with the ground truth device.

\fakepar{Outdoor environment} We evaluate the system outdoors in an environment with unrelated movements, such as people walking and vehicle traffic. We present the results in Figure~\ref{fig:breathing_comparison_outdoor_10cm}. Although the raw values from the system are noisy, they still follow the general trend of the raw force values from the reference device. We post-process the system data using outlier removal and smoothing filters and compensate for the delay using the cross-correlation lag, as illustrated in Figure~\ref{fig:breathing_comparison_outdoor_10cm}. With additional processing and the application of more complex feature extraction through machine learning methods, we anticipate that the system will perform well in outdoor environments. Nonetheless, \system\space operates less reliably and at shorter distances in an outdoor environment.
\begin{figure}[t!]
    \centering
    \includegraphics[width=0.42\textwidth]{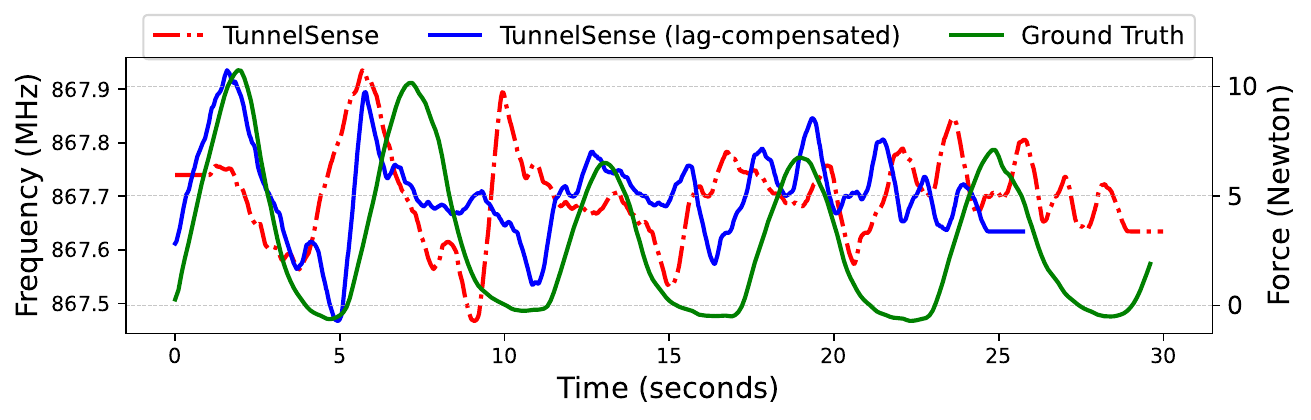}
    \vspace{-4mm}
    \caption{The frequency drifts of radio waves are noisy in an outdoor environment. Nonetheless, at close distances of up to \SI{10}{\centi\meter}, the patterns in frequency drifts corresponding to breathing of subject are still noticeable.}
        \vspace{-4mm}
    \label{fig:breathing_comparison_outdoor_10cm}
\end{figure}
\begin{figure}[t!]
\centering
\includegraphics[width=0.36\textwidth]{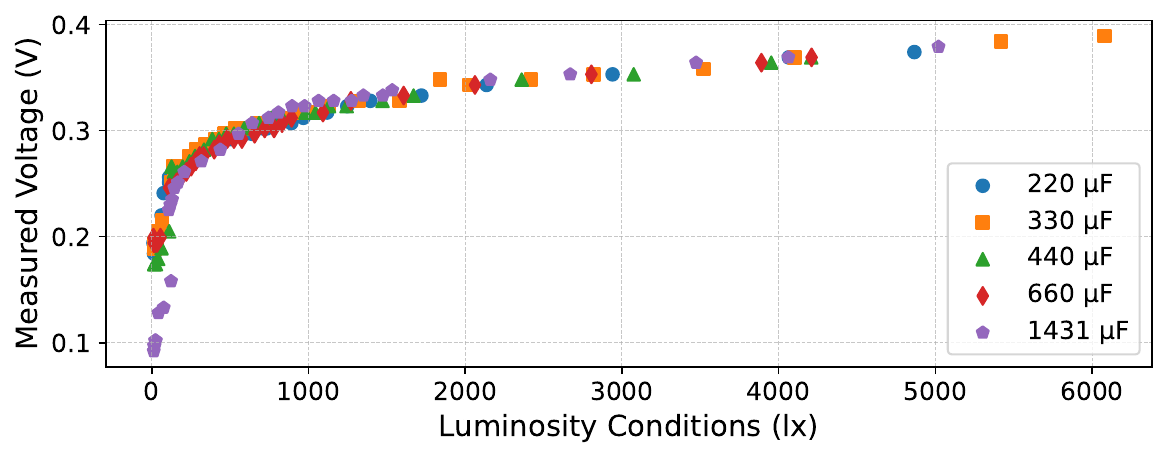}
\vspace{-4mm}
\caption{Peak charge stored (peak voltage) across low-leakage tantalum capacitors under diverse luminosity~(lx) conditions using a photodiode. The stored charge is sufficient to operate \system\space for a short period of time.}
\label{fig:peak_voltage_capacitors}
\vspace{-2mm}
\end{figure}
\fakepar{Operation with harvested energy} To evaluate \system\space operation on energy harvested, we employ a photodiode~\cite{ams_osram} and store the gathered energy in the form of charge across low-leakage tantalum capacitors~\cite{kyoceraavxSeriesKYOCERA}. Figure~\ref{fig:peak_voltage_capacitors} shows the amount of charge (peak voltage) stored across five different capacitors, i.e., 220$\mu$F, 330$\mu$F, 440$\mu$F, 660$\mu$F, and 1431$\mu$F, under different luminosity conditions. We also exhibit the active time (ON state) of the 1N3712 tunnel diode along with its region of operation (in terms of voltage) powered using a 1431$\mu$F capacitor in Table~\ref{tab:TDO_active_time_1431uF}. This indicates the ability of the \system\space tag to operate on minuscule amounts of energy even without using an EHIC for harvesting energy.
\begin{table}[t!]
\centering
\begin{tabular}{|c|c|c|}
\hline
\begin{tabular}[c]{@{}c@{}}Tunnel Diode\\ Part Number\end{tabular} & \begin{tabular}[c]{@{}c@{}}Active Time \\ (in millisecond)\end{tabular} & \begin{tabular}[c]{@{}c@{}}Region of Active Transmission\\ (in millivolts)\end{tabular} \\ \hline 
1N3712~\cite{1N3712} & 51 & 75-197 \\ \hline 
\end{tabular}
\caption{Active time and region of operation (in millivolts) of \system\space tag (peak voltage) on a \SI{1431}{\micro\farad} capacitor.}
\label{tab:TDO_active_time_1431uF}
\vspace{-8mm}
\end{table}

\section{Discussion and Future Work}

 \fakepar{Applications} Sensing breathing rate is essential for diagnosing several health conditions. It helps to understand sleep patterns such as sleep apnea~\cite{Sleepapnea22}. Such scenarios require continuous monitoring of the user's breathing patterns. The \system\space tag can be located close to the user, affixed on a wall, or kept near a desk. Typically, these locations are within the sensing range of the tag, which is tens of centimeters. On the other hand, the receiver device can be farther away from the tag and be located in the same room or another room due to the sufficient communication range. The sensed heart rate could be communicated to healthcare providers or caregivers in real-time, ensuring appropriately timed intervention.

Tracking motion enables numerous other application scenarios. It can help track body movements to infer gestures for controlling computing devices. The sensitivity of the \system\space can  also allow intrusion detection to secure perimeters. 
 
\fakepar{Form-factor} The simplicity of the circuit, consisting of just a handful of components, allows us to realize the tag in novel form factors. Designing tags in forms like stickers or other novel forms is possible. The low power consumption means bulky batteries are unnecessary, and a tiny capacitor may suffice. This presents an interesting possibility for enabling widespread deployment of the \system\space system.

\fakepar{Limitations} \system\space detects breathing rate reliably in less noisy environments. We have seen that the sensing range decreases substantially in noisy and outdoor environments. Nonetheless, a pattern corresponding to the person's breathing rate is still visible even in these environments. We will work to improve this limitation by exploring the use of directional antennas and appropriate machine-learning methods to infer breathing rates in such  complex environments.

Tunnel diodes have considerable untapped potential for enabling energy-efficient communication~\cite{varshney2022judo} and sensing~\cite{tunnelifi}. However, tunnel diodes are commercially obsolete today. Consequently, procuring tunnel diodes is a challenge. We hope this work, along with other recent efforts, will motivate efforts in commercializing tunnel diodes fabrication at scale.

Finally, we acknowledge that the turn-on time for the tag without EHIC for a small capacitor may not be sufficient to track breathing. We will explore increasing the duration by enhancing the storage capacity (capacitance) or by increasing the number of photodiodes or similar energy sources.

\section{Conclusion}
We introduced \system, a novel, low-power system capable of tracking breathing rate in a noncontact manner within a \SI{30}{\centi\meter} range of the subject from the tag. Enabled by the low-power characteristics of tunnel diodes, it operates at voltages as low as \SI{70}{\milli\volt} and power consumption under \SI{100}{\micro\watt}. The \system\space system does not require complex self-interference cancellation mechanisms at the receiver, allowing commodity transceivers to sense \system\space signals. This reduces deployment costs and complexity, enabling the ubiquitous deployment of the \system\space  system.
\section*{Acknowledgement}
We thank the anonymous reviewers for their insightful comments. We acknowledge Pramuka Medaranga and Spanddhana Sara for their  assistance in conducting of some experiments. This work is primarily funded by a grant from the Advanced Research and Technology Innovation Centre (ARTIC)~(A-8000976-00-00), and also partially supported by a startup grant~(A-8000277-00-00), and MoE Tier 1 grant~(A-8001661-00-00) hosted at the National University of Singapore.


\bibliographystyle{plain}
\bibliography{references}
\end{document}